\documentclass{mem}
\usepackage{natbib}\usepackage{txfonts}\usepackage{balance}
\usepackage{graphicx}
\usepackage[a4paper,breaklinks,dvipdfm]{hyperref}
\usepackage{xcolor}
\idline{75}{282}
\begin{document}
\def\teff{$T\rm_{eff }$}
\def\kms{$\mathrm {km s}^{-1}$}

\title{Lithium abundances in globular clusters}

   \subtitle{}

\author{N. \,Sanna\inst{1},  
        E. \,Franciosini\inst{1}, 
        E. \,Pancino\inst{1,2},
        A. \,Mucciarelli\inst{3}}

\institute{
Istituto Nazionale di Astrofisica --
Osservatorio Astrofisico di Arcetri, Largo E. Fermi 5, 50125 Firenze, Italy
\email{nicoletta.sanna@inaf.it}
\and
Space Science Data Center,  ASI, Via del Politecnico snc, 00133 Roma, Italy
\and
Dipartimento di Fisica e Astronomia, Universit\'a degli Studi di Bologna, Via Gobetti 93/2, 40129 Bologna, Italy
}

\authorrunning{Sanna}

\titlerunning{Lithium in GCs}

\abstract{Lithium is created during the Big Bang nucleosynthesis and it is
destroyed in stellar interiors at relatively low
temperatures.  However, it should be preserved in the stellar envelopes of
unevolved stars and  progressively diluted during mixing
processes.  In particular, after the first dredge-up along the RGB, lithium
should be completely destroyed, but this is not what we observe today in
globular clusters.  This element allows to test stellar evolutionary models, as
well as different types of polluters for second population stars in the multiple
population scenarios.  Due to the difficulty in the measurement of the small
available lithium line, few GCs have been studied in details so
far. Literature results are not homogeneous  for what concerns
type of stars, sample sizes, and chemical analysis methods. The Gaia-ESO survey
allows us to study the largest sample of GCs stars (about 2000,
both dwarfs and giants) for which the lithium has been analysed homogeneously. 
\keywords{Stars: abundances -- Galaxy: globular clusters }} 

\maketitle{}

\section{Introduction}

Lithium (Li) is one of the elements created during the Big Bang nucleosynthesis
and it is destroyed in the stellar interiors at relatively low temperatures 
($\sim2.5$x$10^6K$). However, it should be preserved in the stellar envelopes
of unevolved stars and diluted by diffusion and mixing processes. 

\begin{table*}[htbp!]
\caption{\footnotesize GCs included in GES, with the main
evolutionary phases covered by the members sample.}
\label{table:gc}
\begin{center}
\begin{tabular}{|ll|ll|ll|}
\hline \hline
Cluster&Phase&Cluster&Phase&Cluster&Phase\\
\hline
M12&RGB, AGB&M15&RGB, AGB&M2&RGB, AGB\\
NGC104&MS, RGB, AGB&NGC362&RGB, AGB&NGC1261&RGB, AGB\\
NGC1851&RGB, AGB&NGC1904&RGB, AGB&NGC2808&RGB, AGB\\
NGC4372&RGB, AGB&NGC4590&RGB, AGB&NGC4833&RGB, AGB\\
NGC5927&RGB, AGB&NGC6553&RGB, AGB&NGC6752&MS, RGB, AGB\\
\hline
\end{tabular}
\end{center}
\end{table*}

The study of Li abundance in Globular Clusters (hereafter
GCs), among the oldest objects in the Galaxy, allows to investigate the cosmic
lithium problem, as well as to  give important information on
stellar evolution and mixing processes and on the formation of
multiple populations. Few GCs have been investigated so far in detail, and
different evolutionary phases have  been considered, but the emeriging scenario
is complex. In general, the dwarf stars share the same lithium abundance,
$A(Li)\sim2.2$ dex as found by \citet{spitea, spiteb}, forming
the so called Spite plateau.  During the sub-giant branch phase
diffusion processes dilute Li, as well as
mixing during the first dredge-up process in the red giant
branch (hereafter RGB). However, \citet{mucciarelli12} found that stars above
the first dredge-up and below the RGB bump form another plateau
at about $A(Li)\simeq1.0$ dex, while after the bump Li is
diluted again. The above framework is further complicated by
the phenomenon of multiple populations in globular clusters. In fact, in some
clusters (e. g., NGC~362) first and second population stars
have the same Li abundance, while in others (e. g., NGC~6752)
they do not, and in the case of $\omega$ Cen, both behaviors are observed at the
same time \citep[see][and reference therein, for details]{mucciarelli18}.

Moreover, 14 Li-rich stars have been discovered so far in 12
different GCs \citep{mucciarelli19}. The majority are RGB
stars, with only two main sequence (hereafter MS) stars and two
that belong to the asymptotic giant branch (hereafter AGB). Different
enrichment scenarios have been proposed to explain their
existance, such as engulfment of sub-stellar systems,
self-enrichment based on the Cameron-Fowler mechanism \citep{cameronfowler} and
mass-transfer, but a consensus has not been reached yet. 

The existence of stars enhanced in lithium has also important
implications in terms of the formation of multiple populations in GCs. Being so
fragile, Li is a powerful tracer of the physical conditions of the polluting
source, and poses strict limits to the temperatures involved in the process.

\section{Data sets and analysis}

This work is based on spectra gathered by the Gaia-ESO survey
\citep[hereafter GES]{gilmore12,randich13} and on its recommended parameters and
abundances. The spectra have been acquired using FLAMES-GIRAFFE
\citep{pasquini} and FLAMES-UVES \citep{dekker00}, combining new observations
and ESO archival data obtained with the GES instrumental
setups. The reduction of the GIRAFFE and UVES spectra was performed as
described in \citet{jackson15} and \citet{smiljanic}, respectively. The
atmospheric parameters and  abundances were determined with the GES
multiple-pipeline strategy
\citep{recioblanco14,smiljanic,pancino17}. The lithium
abundance has been derived using the Li 6707 \AA~doublet, taking into account
the blend with the neighboring Fe line. Within GES, 15 GCs were observed as part
of the survey calibration strategy \citep{pancino17,pancino17a}, covering a
range of metallicities and global properties. We list them in
Table~\ref{table:gc} together with the evolutionary stages sampled by the member
stars. In particular, RGB stars have been observed
in all GCs, while for two GCs MS stars from the archive sample
have been analyzed as well: this represents the largest sample homogeneously
analyzed ever used to spectroscopically investigate Lithium in GCs. 
Of the about 2500 stars analyzed in GES, we selected almost
2000 bona-fide members using the GES [Fe/H] and radial velocity determinations,
following \citet{pancino17a}.

\begin{figure*}[htbp!]
\resizebox{\hsize}{!}{\includegraphics[clip=true]{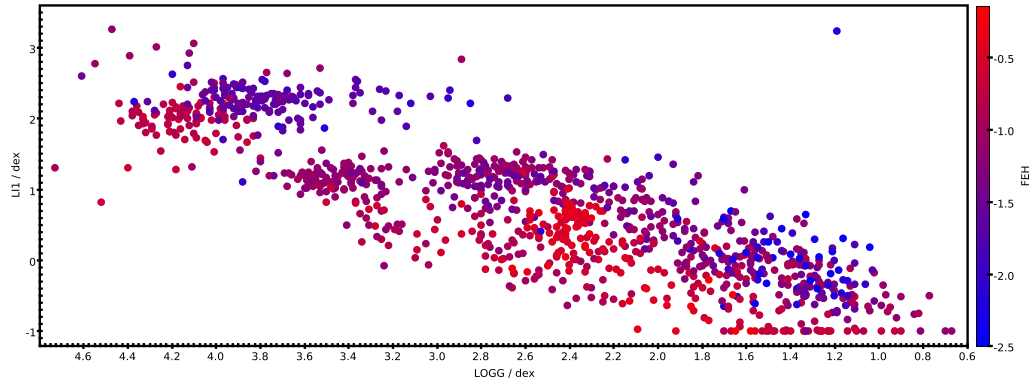}}
\caption{\footnotesize Li abundance as a function of surface
gravity for the member stars of the 15 {GCs included in GES},
color-coded by metallicity.  The Spite plateau is well defined, at
{A(Li)~$\simeq 2.2$ dex}, as well as the plateau
defined by stars above the first dredge-up and below the bump
along the RGB, at A(Li)~$\simeq 1.1$ dex. The evolutionary
phases of the stars included in the sample are clearly identified.} 
\label{li}
\end{figure*}

\section{Discussion}

Our knowledge about Li in GCs is based on inhomogeneous
studies, focused on a few clusters, investigated with varying sample sizes, and
analyzed with different methods. For example, in the case of NGC~6752 only 9 MS
stars have been investigated \citep{pasquini05}, while $\simeq 350$ stars (from
below the MS turnoff up to the bright end of the RGB) have been studied in the
case of NGC~6397 \citep{lind09}. 

GES provides an unprecedented data set, that allows us to
effectively attack the lithium problem in GCs, thanks to the very high quality
of the spectra and the strategy used to homogenize the abundances.
Typical uncertainties in the abundances of metal-poor stars
amount to about 0.1~dex or less \citep{pancino17a}. In this respect, we can also
take advantage of the relatively high number of GCs included in the survey and
on their different properties, covering a range of global GC parameters like
metallicity, total mass, concentration, and so on.

Figure~\ref{li} shows the Li abundances of the $\simeq$2000
selected member stars in the 15 GCs, determined by GES, as a function of surface
gravity (log$g$). The Spite plateau defined by dwarf stars
appears well defined at A(Li)~$\simeq 2.2$ dex, as well as
the plateau defined by stars above the first dedge-up and below
the RGB bump, at A(Li)~$\simeq 1.1$ dex. The RGB stars brighter than the
RGB bump show the expected behaviour, whit Lithium
progressively diluted due to mixing processes. The stars
in Figure~\ref{li} are color-coded by metallicity
showing several effects that are worth further investigation.

\subsection{Li-rich stars} 

As already mentioned above, 14 Li-rich stars in GCs are known
in the literature, and their origin mechanism is still not well
understood. Two of them are included in the GES sample
presented here: the RGB bump star discovered in  NGC~362 by
\citet{dorazi} and the bright RGB or AGB star discovered in
NGC~4590 by \citet{ruchti} and \citet{kirby}. 

\begin{figure}[]
\resizebox{\hsize}{!}{\includegraphics[clip=true]{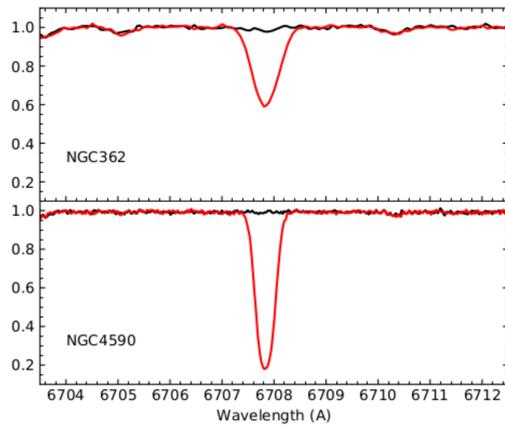}}
\caption{\footnotesize Top panel: comparison between the
spectrum of the Li-rich star in NGC~362 (red line) and a reference member star
with similar atmospheric parameters. Bottom panel: a similar comparison in the
case of the Li-rich star in NGC~4590.}
\label{li_rich}
\end{figure}

Here we take the occasion and use these stars to demonstrate
the exquisite quality of the GES spectra.  Figure~\ref{li_rich} shows the
comparison between the {spectra of the two Li-rich stars in
NGC~362 and NGC~4590 observed by GES, compared with Li-normal member stars of
the two GCs, having similar atmospheric parameters. The typical Li enhancement
with respect to similar cluster stars is of at least one order of magnitude. The
two Li-rich stars are well visible also in Figure~\ref{li}, where they display
A(Li)$\simeq$2.7~dex in the case of NGC~362 and A(Li)$\simeq$3.3~dex in the case
of NGC~4590. These preliminary abundance estimates are not yet corrected for
NLTE effects.}

\section{Conclusions}

Using prelimnary abundances from the Gaia-ESO Survey, we have
demonstrated the power of the GES dataset for a global understanding of lithium
in GCs. Almost 2000 member stars at different evolutionary stages were analyzed
homogeneously, belonging to 15 GCs with different metallicity, mass,
concentration, orbit, and selected to cover well the space of GC global
properties as calibrators for GES. This sample offers the possibility  to
investigate stars with different properties, such as temperature, gravity and
metallicity, in the largest sample homogeneously analyzed so far. The exquisite
quality of the spectra allows us to well determine the Li abundance and identify
rare Li-rich stars in this kind of systems.
The GES dataset will enable a full statistical investigation of
the Li behaviour in GCs, and it will undoubtedly provide an invaluable reference for further
theoretical studies, allowing for a deeper understanding of stellar evolution
and providing hopefully new insights on the multiple stellar population problem
in GCs.


\bibliographystyle{aa}

\begin{thebibliography}{}
\bibitem[Cameron \& Fowler(1971)]{cameronfowler}
Cameron, A. G. W. \& Fowler, W. A. 1971, ApJ, 164, 111

\bibitem[Dekker et al.(2000)]{dekker00}
Dekker, H., D'Odorico, S., Kaufer, A. et al. 2000, SPIE, 4008, 534

\bibitem[D'Orazi et al.(2015)]{dorazi}
D'Orazi, V., Gratton, R. G., Angelou, G. C. et al. 2015, ApJ, 801, L32

\bibitem[Gilmore et al.(2012)]{gilmore12}
Gilmore, G., Randich, S., Asplund, M. et al. 2012, The Messenger, 147, 25

\bibitem[Jackson et al.(2015)]{jackson15}
Jackson, R. J., Jeffires, R. D., Lewis, J. et al. 2015, A\&A, 580, 75

\bibitem[Kirby et al.(2016)]{kirby}
Kirby, E. N., Guhathakurta, P., Zhang, A. J. et al. 2016, ApJ, 819, 135

\bibitem[Lind et al.(2009)]{lind09}
Lind, K., Primas, F., Charbonnel, C. et al. 2009, A\&A, 503, 545

\bibitem[Mucciarelli et al.(2012)]{mucciarelli12}
Mucciarelli, A., Salaris, M., Bonifacio, P. 2012, MNRAS, 419, 2195

\bibitem[Mucciarelli et al.(2018)]{mucciarelli18} 
Mucciarelli, A., Salaris, M., Monaco, L. et al. 2018, A\&A, 618, 134

\bibitem[Mucciarelli et al.(2019)]{mucciarelli19} 
Mucciarelli, A., Monaco, L., Bonifacio, P. et al. 2019, A\&A, 623, 55

\bibitem[Pancino et al.(2017a)]{pancino17} Pancino, E., Lardo, C., 
Altavilla, G., et al.\ 2017, \aap, 598, A5

\bibitem[Pancino et al.(2017b)]{pancino17a} Pancino, E., Romano, D., 
Tang, B., et al.\ 2017, \aap, 601, A112

\bibitem[Pasquini et al.(2000)]{pasquini} 
Pasquini, L., Avila, G., Allaert, E. et al. 2000, SPIE, 4008, 129

\bibitem[Pasquini et al.(2005)]{pasquini05}
Pasquini, L., Bonifacio, P., Molaro, P. et al. 2005, A\&A, 441, 549

\bibitem[Randich et al.(2013)]{randich13} 
Randich, S., Gilmore, G. \& the Gaia-ESO Consortium 2013, The Messenger, 154, 47

\bibitem[Recio-Blanco et al.(2014)]{recioblanco14}
Recio-Blanco, A. de Laverny, P., Kordopatis, G. et al. 2014, A\&A, 567, 5

\bibitem[Ruchti et al.(2011)]{ruchti}
Ruchti, G. R., Fulbright, J. P., Wyse, R. F. G. et al. 2011, ApJ, 743, 107

\bibitem[Smiljanic et al.(2014)]{smiljanic}
Smiljanic, R., Korn, A. J., Bergemann, M. et al. 2014, A\&A, 570, 122

\bibitem[Spite \& Spite(1982a)]{spitea}
Spite, F., Spite, M. 1982a, A\&A, 115, 357

\bibitem[Spite \& Spite (1982b)]{spiteb}
Spite, M., Spite, F. 1982a, Nature, 297, 483

\end{thebibliography}

\end{document}